\documentclass[12pt]{article}
\usepackage{graphicx}

\usepackage{latexsym} 
\usepackage{amssymb}  
\usepackage{multirow}

\def\beq{\begin{equation}}
\def\eeq#1{\label{#1}\end{equation}}
\def\eeqn{\end{equation}}

\def\beqa{\begin{eqnarray}}
\def\eeqa#1{\label{#1}\end{eqnarray}}
\def\eeqan{\end{eqnarray}}

\let\bar=\overbar

\def\Dslash{\not{\hbox{\kern-4pt $D$}}}
\def\dslash{\not{\hbox{\kern-2pt $\del$}}}

\def\msb{{\bar{\ssstyle M \kern -1pt S}}}

\def\Title#1{\begin{center} {\Large {\bf #1} } \end{center}}

\begin{document}

\Title{Color superconducting quark matter\\[1ex] in compact stars}

\bigskip\bigskip


\begin{raggedright}  

{\it Mark Alford\index{Alford, M.}\\
Physics and Astronomy Department\\
Glasgow University \\
Glasgow G12 8QQ, UK }
\bigskip\bigskip
\end{raggedright}

\section{Introduction}
\label{sec:int}

Matter just above nuclear density presents special challenges,
both observational and theoretical.
It is the densest state of which we have evidence in nature,
but in terrestrial laboratories it can only be produced 
fleetingly and at high temperature over small volumes. 
We must therefore rely on
astrophysical observations of compact stars to gather 
more indirect experimental information.
Theoretically, it falls between the domain of nuclear physics, where
models have been tested against scattering data, and the 
asymptotically high energy
domain, where the relevant theory, QCD, becomes tractable.

We expect that somewhat above nuclear density, the nucleons will
overlap so much as to lose their separate identities,
and merge into quark matter.
In this talk I will review some of theoretical
expectations and speculations about quark matter, focusing
on the phenomenon of quark pair condensation (color superconductivity).

\index{asymptotic freedom}
\index{Fermi surface}
Since QCD is asymptotically free, one expects that at high enough
densities and low temperatures, matter will consist of a Fermi sea of
quarks, and the ones at the Fermi surface will be almost free.
The residual gluon-mediated interaction is attractive 
in the color $\bar {\bf 3}$ channel, so BCS quark pair condensation
will take place, breaking the $SU(3)$ color gauge symmetry.
\index{BCS mechanism}
This was first appreciated in the 1970s, and revived more recently
\cite{Barrois,BarroisPhD,BailinLove,ARW2,RappETC}.
It is discussed in more detail in the review articles \cite{Reviews}.
The quark pairs play the same role here as the Higgs particle
does in the standard model: the color-superconducting phase
can be thought of as the Higgsed (as opposed to confined)
phase of QCD.

\index{order parameters}
It is important to remember from the outset
that the breaking of a gauge symmetry
cannot be characterized by a gauge-invariant local order parameter
which vanishes on one side of a phase boundary. The superconducting
phase can be characterized rigorously only by its global symmetries.
In electromagnetism there is a non-local order parameter, the 
mass of the magnetic photons, that corresponds physically to the Meissner
effect and distinguishes the free phase from the superconducting one.
In QCD there is no free phase: even without pairing the gluons are not
states in the spectrum.  No order parameter distinguishes the Higgsed
phase from a confined phase or a plasma, so we have to look at the
global symmetries.

\section{Patterns of color superconductivity}
\label{sec:flavor}

In the real world there are two light quark flavors, the up 
($u$) and down ($d$), with
masses $\lesssim 10~{\rm MeV}$, and a medium-weight flavor, the strange 
($s$) quark, with mass $\sim 100~{\rm MeV}$.
It is convenient to treat the $u$ and $d$ as massless,
and study the effect of varying the $s$ quark mass between
zero and infinity.

\subsection{Three flavors: color-flavor locking}
In the three flavor case, $m_s=m_u=m_d=0$, the structure of the
quark pair condensate is particularly simple and elegant, because
the number of flavors and colors is equal.
The favored condensation pattern is ``color-flavor locking'',
\beq
\mbox{CFL~phase:}\quad
\Delta^{\alpha\beta}_{ij} = \langle q^\alpha_i q^\beta_j \>^{\phantom\dagger}_{1PI}
\propto C \gamma_5 [ (\kappa+1)\delta^\alpha_i\delta^\beta_j + (\kappa-1) \delta^\alpha_j\delta^\beta_i],
\label{3flav:vev}
\eeqn
where color indices $\alpha,\beta$ and flavor indices 
$i,j$ all run from 1 to 3, Dirac indices are suppressed,
and $C$ is the Dirac charge-conjugation matrix.
The term multiplied by $\kappa$ corresponds to pairing in the
$({\bf 6}_S,{\bf 6}_S)$, which
although not highly favored energetically 
breaks no additional symmetries and so
$\kappa$ is in general small but not zero 
\cite{ARW3,PisarskiCFL,SchaeferPatterns}.

Eq.~(\ref{3flav:vev}) exhibits the color-flavor locking
property of this ground state. The Kronecker deltas connect
color indices to flavor indices, so that the VEV is not
invariant under color rotations, nor under flavor rotations,
but only under simultaneous, equal and opposite, color and flavor
rotations. Since color is only a vector symmetry, this
VEV is only invariant under vector flavor rotations, and
breaks chiral symmetry.

The pattern of symmetry breaking is therefore (with gauge symmetries
in square brackets)
\beq
[SU(3)_{\rm color}]
 \times \underbrace{SU(3)_L \times SU(3)_R}_{\displaystyle\supset [U(1)_Q]}
 \times U(1)_B 
\longrightarrow \underbrace{SU(3)_{C+L+R}}_{\displaystyle\supset [U(1)_{{\tilde Q}}]}
 \times \mathbb{Z}_2 
\eeqn
Note that electromagnetism is not a separate
symmetry, but corresponds to gauging one of the flavor generators.

This pattern of condensation has many interesting features.
(1) The color gauge group is completely broken, so all eight gluons
become massive. This ensures that there are no infrared divergences
associated with gluon propagators.
(2) All the quark modes are gapped. The nine quasiquarks 
(three colors times three flavors) fall into a smaller-gap octet
and a larger-gap singlet of the unbroken global $SU(3)$.
(3) Electromagnetism is replaced by a ``rotated electromagnetism'',
namely a linear combination ${\tilde Q}$ of the original photon and one of the gluon.
\index{rotated electromagnetism}
(4) Two global symmetries are broken,
\index{chiral symmetry breaking}
the chiral symmetry and baryon number, so there are two 
gauge-invariant order parameters
that distinguish the CFL phase from the QGP,
with corresponding Goldstone bosons which are long-wavelength
disturbances of the order parameter. 
(5)
Quark-hadron continuity. It is striking that the symmetries of the
3-flavor CFL phase are the same as those one might expect for 3-flavor
hypernuclear matter \cite{SW-cont}.  This means the spectrum may
evolve continuously from  hypernuclear matter to
the CFL phase of quark matter---there need be no phase transition.

\subsection{Two flavors}

If the strange quark is heavy enough to be ignored, then
the up and down quarks
pair in the color ${\bf \bar 3}$ flavor singlet
channel, a pattern that we call the two-flavor superconducting (2SC)
phase,
\beq
\label{2flav:vev}
\mbox{2SC~phase:}\quad
\Delta^{\alpha\beta}_{ij} =
\langle q^\alpha_i  q^\beta_j \>^{\phantom\dagger}_{1PI} 
  \propto C \gamma_5\varepsilon_{ij}\varepsilon^{\alpha\beta 3},
\eeqn
where color indices $\alpha,\beta$ run from 1 to 3, flavor indices 
$i,j$ run from 1 to 2. 
Four-fermion interaction calculations agree on the
magnitude of $\Delta$: around $100~{\rm MeV}$. This is found to be
roughly independent of the cutoff, although the chemical potential
at which it is attained is not.
Such calculations are based on
calibrating the coupling to give a chiral condensate of around
$400~{\rm MeV}$ at zero density, 
and turning $\mu$ up to look for the maximum gap.

As with any spontaneous symmetry breaking, one of the degenerate
ground states is arbitrarily selected.
In this case, quarks of the first two colors
(red and green) participate in pairing, while the third color
(blue) does not.
The ground state is invariant under an $SU(2)$
subgroup of the color rotations that mixes
red and green, but the blue quarks are singled out as different.
The pattern of symmetry breaking is therefore 
\beq
\label{2flav:syms}
\begin{array}{rl}
& [SU(3)_{\rm color}]\times [U(1)_Q]
 \times SU(2)_L \times SU(2)_R \\
\longrightarrow & 
 [SU(2)_{\rm color}]\times [U(1)_{{\tilde Q}}]
 \times SU(2)_L \times SU(2)_R \\
\end{array}
\eeqn
The features of this pattern of condensation are
(1)
The color gauge group is broken down to $SU(2)$, so five of the gluons
will become massive, with masses of order $g\mu$.
The remaining three gluons are associated with an
unbroken $SU(2)$ red-green gauge symmetry, whose confinement 
distance scale rises exponentially with density \cite{SU2unbroken}.
(2)
The red and green quark modes acquire a gap $\Delta$.
There is no gap for the blue quarks in this ansatz, and it is an
interesting question whether they find some other channel in which to pair.
The available attractive channels appear to be weak
so the gap will be much smaller, perhaps
in the {\rm keV}\ range \cite{ARW2}. We will ignore such pairing here.
(3) A rotated electromagnetism (``${\tilde Q}$'')
survives unbroken. It is a combination
of the original photon and one of the gluons.
\index{rotated electromagnetism}
(4) No global symmetries are broken,
although additional condensates that break chirality have been 
suggested \cite{Berges},
so the 2SC phase has the same global
symmetries as the quark-gluon plasma (QGP).

\section{Two massless + one massive quark flavors}

A nonzero strange quark mass explicitly breaks the 
flavor $SU(3)_L\times SU(3)_R$ symmetry down to
$SU(2)_L\times SU(2)_R$. 
If the strange quark is heavy enough then it will decouple,
and 2SC pairing will occur.
For a sufficiently small strange 
quark mass we expect a reduced form of color-flavor 
locking in which an $SU(2)$ subgroup of $SU(3)_{\rm color}$
locks to isospin, causing chiral symmetry breaking and
leaving a global $SU(2)_{{\rm color}+V}$ group unbroken.

As $m_s$ is increased from zero to infinity, there has to be some 
critical value at which the strange quark decouples,
color and flavor rotations are unlocked, and 
the full $SU(2)_L \times SU(2)_R$ symmetry is restored \cite{2+1flav}.
On the way, however, other interesting phenomena may occur.
Where differing masses or chemical potentials
obstruct the pairing of one species of quark with another
(eg in the 2SC+s phase)
we expect regions of crystalline superconductivity \cite{OurLOFF},
which grow larger at high chemical potential where forward
scattering dominates the quark-quark interactions
\cite{Leibovich:2001xr}.
This could lead to interesting phenomena such as glitches in
quark stars \cite{OurLOFF}.
It is also possible that the CFL condensate responds to the
strange quark mass by rotating in a direction that reduces the
strangeness content. This corresponds to a condensation
of $K_0$ mesons \cite{BedaqueSchaefer} yielding a ``CFL-$K^0$''
phase.

\begin{table}[htb]
\newlength{\lena} \settowidth{\lena}{charge}
\newlength{\lenb} \settowidth{\lenb}{electromag}
\newlength{\lenc} \settowidth{\lenc}{number}
\begin{tabular}{l@{}rccccc}
\hline
symmetry: & gauged & \parbox{\lenc}{baryon\\[-0.3ex] number}
 & \parbox{\lena}{hyper-\\[-0.3ex] charge}
 & isospin & chiral & axial $U(1)$\\
\hline
broken by: & & & weak & 
  \parbox{\lenb}{electromag \\[-0.5ex]  $m_u\neq m_d$} 
  & $m_{u,d}\neq 0$ & instanton \\
\hline
QGP & $[SU(3)\times U(1)_Q]$
  & $U(1)_B$ & $U(1)_Y$ & $SU(2)_V$ & $SU(2)_A$ & $U(1)_A$ \\
2SC & $[SU(2)\times U(1)_{\tilde Q}]$
  & $U(1)_B$ & $U(1)_Y$ & $SU(2)_V$ & $SU(2)_A$ & $U(1)_A$ \\
CFL &  $[U(1)_{{\tilde Q}}] $ 
  &  \{1\} & $U(1)_Y$ & $SU(2)_V$ & \{3\} & \{1\}\\
CFL-$K^0$ &   $[U(1)_{{\tilde Q}}]$ 
  &  \{1\} & \{1\} & \{2\} & \{3\} & \{1\}\\
\hline
\end{tabular}
\caption{Symmetry breaking of the high-density QCD
phases in various approximations. For broken
symmetries, the number of Goldstone bosons is given in curly brackets.
Gauged symmetries are in square brackets.
The strongest explicit breaking of each of the global symmetries
is given: only baryon number is a global symmetry of the real world.
The QGP is assumed to have the symmetries of the Lagrangian.
}
\label{tab:2+1sym}
\end{table}

\begin{figure}[htb]
\includegraphics[width=0.9\textwidth]{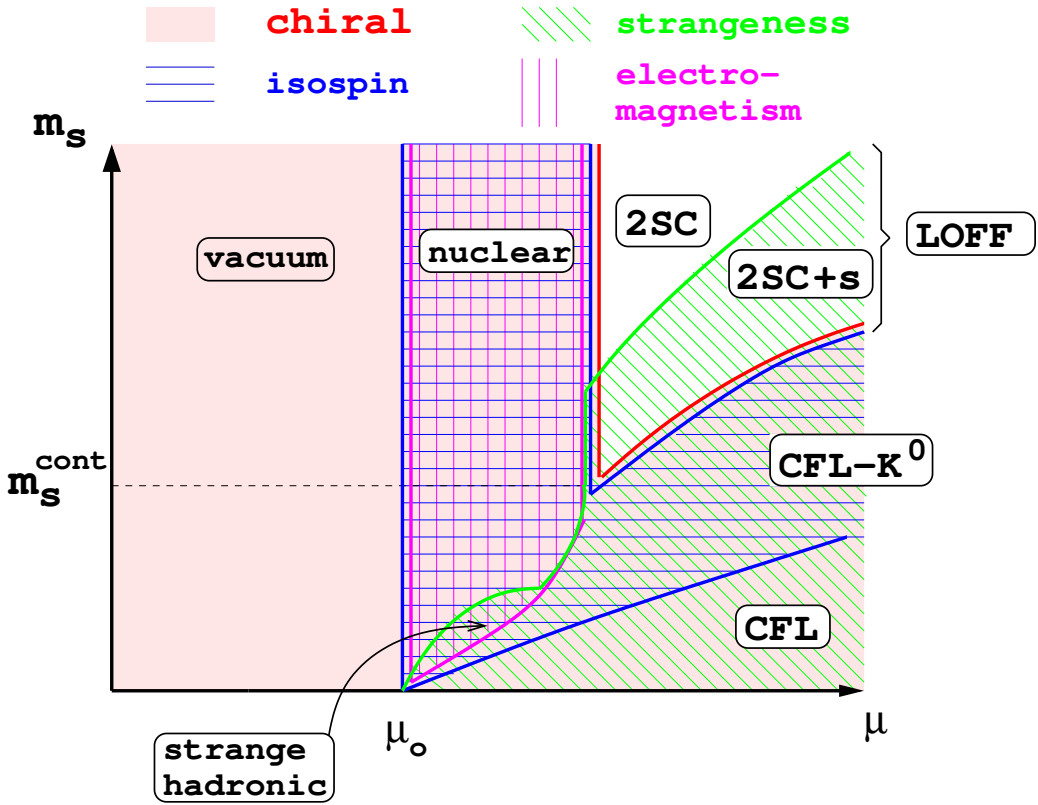}
\caption{Possible phase diagram for 2+1 flavor QCD at $T=0$.
In the blue (horizontal) shaded area, isospin is broken.
In the magenta (vertical) shaded area, electromagnetism
is broken. In the green (top-left to bottom-right) shaded
area, strangeness is broken. Chiral symmetry is broken
in the whole densely (pink) shaded area, i.e.~everywhere
except in the wedge at top right. Small print: most of these
symmetries are approximate, and potentially important
physics has been ignored---see the text.
}
\label{fig:qcd2+1flav}
\end{figure}

The CFL-$K^0$\ phase is expected to occur if the strange quark mass is
large enough, and the up and down are light enough.  Calculations at
asymptotically high density, extrapolated back to around nuclear
density, indicate that it is favored for quark matter at a few
times nuclear density \cite{KaplanReddy}.
This estimate ignores terms that split the gaps corresponding
to different flavor pairings \cite{2+1flav} and
$U(1)_A$-breaking terms
induced by instantons that are expected to become important in that
regime. In table \ref{tab:2+1sym} we show the symmetries broken
by the various phases. In the real world, the only
relevant true global symmetry is baryon number, and both
CFL and CFL-$K^0$\ break it, so they both have one massless superfluid mode.
However, we expect a number of pseudo-Goldstone bosons arising from
spontaneously broken near-symmetries. These are different in the
CFL and CFL-$K^0$\ phases. It has recently been noted that the breaking of
isospin in the CFL-$K^0$\ phase leads to 2 not 3 pseudo-Goldstone bosons,
\cite{kaonGoldstone}, and that the breaking of
$U(1)_Y$ by the $K^0$-condensate may lead to long-lived axion-type
domain walls \cite{SonDomain}.

In Fig.~\ref{fig:qcd2+1flav} we give one possibility for the
resultant QCD phase diagram in the $\mu$-$m_s$ plane.
The complications relating to the CFL-$K^0$\ phase
have been ignored, and
most of the symmetries shown in the figure are approximate,
so the ``symmetry breaking'' shaded areas actually
correspond to regions where some pseudo-Goldstone
states become very light, and the border lines are crossovers.
The exceptions are electromagnetism, which is a gauge symmetry,
with the mass of the photon as an order parameter, and
baryon number.
Electromagnetism is broken in the magenta (vertical) shaded
region. Baryon number is a combination of electric charge,
isospin, and strangeness, and is broken in all the shaded
regions, leading to an exactly massless superfluid mode.

The low-$\mu$ part of the CFL phase is the region of
quark-hadron continuity \cite{SW-cont}, where the
whole baryon octet self-pairs in an isospin-respecting way
\cite{2+1flav}. The low-$\mu$ part of the CFL-$K^0$\ phase
can also be given a hadronic interpretation, in which
there is $n$-$n$ and $p$-$\Sigma^-$ pairing, which breaks
isospin amd strangeness, but leaves electromagnetism unbroken.

\section{The transition to color superconducting quark matter
in compact stars}
\label{sec:compact}

The only place in the universe where we expect very
high densities and low temperatures is compact stars
(for a recent review, see \cite{HP-nstar}).
These typically have  masses
close to $1.4 M_\odot$, and are believed to have radii of order 10 km.
Their density ranges from around nuclear density near the surface
to higher values further in, although uncertainty about the equation
of state leaves us unsure of the value in the core.


Color superconductivity gives mass to
excitations around the ground state: it opens
up a gap at the quark Fermi surface, and makes the gluons
massive. One would therefore expect its main consequences
to relate to transport properties, such as mean free paths,
conductivities and viscosities.
\index{equation of state}
The influence of color superconductivity on the equation of state
is an $ {\cal O} ((\Delta/\mu)^2)$ (few percent) effect, which is 
probably not phenomenologically
interesting given the existing uncertainty in the equation
of state at the relevant densities.

\subsection{The transition region}
\index{mixed phase}

There are two possibilities for the transition from nuclear matter
to quark matter in a neutron star:
a mixed phase, or a sharp interface. The surface
tension of the interface determines which is favored.

\begin{figure}[htb]
\includegraphics[width=0.9\textwidth]{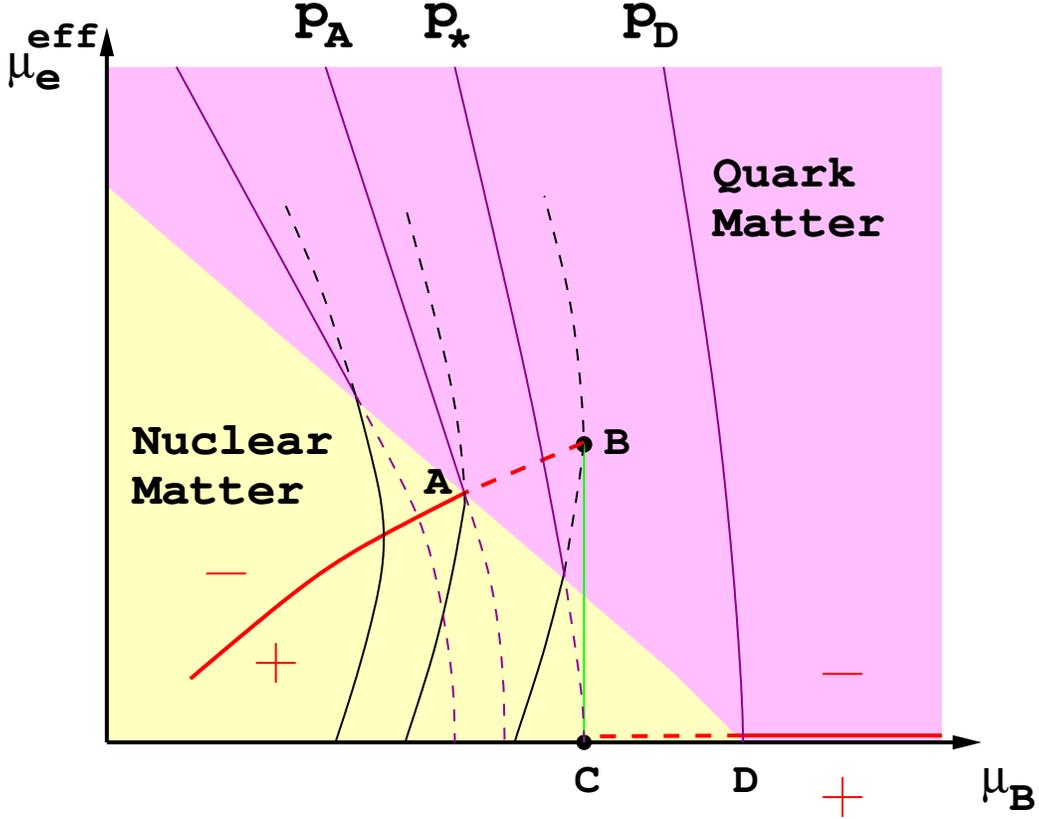}
\caption{A schematic form of the $\mu_B$-$\mu_e$ phase diagram for
nuclear matter and CFL quark matter, ignoring electromagnetism.
For an explanation see the text.}
\label{fig:mixed}
\end{figure}

To be concrete, we will consider the case where the strange quark is light
enough so that quark pairing is always of the CFL type.
Figure \ref{fig:mixed} shows the $\mu_B$-$ \mu^{\rm eff}_e$ phase diagram, 
ignoring electromagnetism.
The lightly (yellow) shaded region is where nuclear matter (NM)
has higher pressure. The darker (magneta) region is where 
quark matter (QM) has higher pressure.
Where they meet is the coexistence line.
The medium solid lines labelled by values of the pressure
are isobars.
Below the coexistence line they are given by the NM equation of state,
above it by the QM equation of state. 
\index{equation of state}

The thick (red) lines are the neutrality lines. Each phase is negatively
charged above its neutrality line and positively charged below it. 
Dotted lines show extensions onto the unfavored sheet (NM above
the coexistence line, QM below it).

The electric charge density is
\beq
Q = - \left.\frac{\partial p}{\partial \mu_e}\right|_{\mu_B} 
\eeqn
so the neutrality line  goes through the right-most extremum
of each isobar, since there the derivative of pressure 
with respect to $\mu_e$ is zero.
For the CFL phase, the neutrality line is $\mu_e=0$
\cite{CFLneutrality}. 

Two possible paths from nuclear to CFL matter as
a function of increasing $\mu$ are shown.
In the absence of electromagnetism and surface tension, 
the favored option is to progress along
the coexistence line from A to D, giving an
overall neutral phase made of appropriate relative volumes
of negatively
charged CFL matter and positively charged nuclear matter.

If, on the other hand, Coulomb and surface energies are
large, then the system
remains on the nuclear neutrality line up to
$B$, where there is a single interface
between nuclear matter at $B$ 
and CFL matter at $C$. This minimal interface,
with its attendant charged boundary layers \cite{interface}, 
occurs between phases
with the same $\mu_e$, $\mu=\mu_B=\mu_C$, and pressure $P_*$.  
The effective chemical potential $ \mu^{\rm eff}_e$
changes across the interface, though, 
as a result of the presence of the electric field.
For more details see Ref.~\cite{interface}.

As yet, not much work has been done on signatures related to these features.
The single interface creates a dramatic density discontinuity in the star:
CFL quark matter at about four times nuclear density floats on
nuclear matter at about twice nuclear density. This may affect the
mass vs.~radius relationship for neutron stars with quark
matter cores.  It may also have qualitative effects on
the gravitational wave profile emitted during the inspiral
and merger of two compact stars of this type. 
The mixed phase has distinctively short neutrino mean free paths,
due to coherent scattering 
\cite{Reddy:2000ad}. Also, the droplets form a crystal lattice
that could pin vortices, leading to glitches.

\subsection{Latent heat}
If there is a first-order phase transition between 
nuclear matter and quark matter, then there is the possibility that
heat could be released when the center of a gravitationally
compressed neutron star converts to quark matter \cite{hypernovae}.

A proper treatment requires the construction
of solutions to the TOV equations, and a comparison of the
masses of the resultant stars. To get an idea of the energy scales
involved, however, one can study the statistical mechanics
of a simplified system consisting
of a lump of nuclear matter, slowly being compressed in a piston.
We assume the system remains at zero temperature throughout.
Since the compression is slow, the system will remain
in equilibrium at all times, which means it can be characterized
by a chemical potential $\mu$ for quark number. This is true
in spite of the fact that the system as a whole has a fixed quark
number: since it is in equilibrium, any subsystem has the same
intensive properties as the whole system, and the rest of
the system acts as a particle reservoir for the subsystem.

Under slow compression of nuclear matter, the pressure $p$ and
chemical potential $\mu$ rise, until we reach a point in phase space
where quark matter with the same chemical potential $\mu_*$ would have
the same pressure $p_*$.
\beq
\label{compact:equilm}
\begin{array}{rclcl}
- p_* &=& \displaystyle\frac{F_{NM}}{V_{NM}} 
  &=& \displaystyle\frac{E_{NM}}{V_{NM}} - \mu_* \frac{N_{NM}}{V_{NM}} \\[3ex]
- p_* &=& \displaystyle\frac{F_{QM}}{V_{QM}} 
  &=&  \displaystyle\frac{E_{QM}}{V_{QM}} - \mu_* \frac{N_{QM}}{V_{QM}} 
\end{array}
\eeqn
Under continued compression, the system contracts
and $\mu$ and $p$ remain constant for a while, as
the nuclear matter is converted to the denser quark matter phase.
When the conversion is complete, further compression causes
$\mu$ and $p$ to start rising again.
From (\ref{compact:equilm}) we see that the quark matter phase
has higher energy density, so the latent heat per unit
volume is negative,
\beq
\label{compact:pervol}
\frac{E_{NM}}{V_{NM}} - \frac{E_{QM}}{V_{QM}} 
  = \mu_* \Bigl( \frac{N_{NM}}{V_{NM}} - \frac{N_{QM}}{V_{QM}} \Bigr)
\eeqn
The relevant quantity, however, is the amount of energy liberated
per quark, since the quark number $N=N_{NM}=N_{QM}$ is constant,
and the two phases have different volumes $V_{NM} > V_{QM}$.
From (\ref{compact:equilm})
the latent heat per quark is
\beq
\label{compact:latH}
\Delta E/N = \frac{E_{NM}}{V_{NM}}\frac{V_{NM}}{N}
- \frac{E_{QM}}{V_{QM}}\frac{V_{QM}}{N} 
= p_* \Bigl( \frac{V_{QM}}{N} - \frac{V_{NM}}{N} \Bigr)
\eeqn
which is again negative. The energy required 
comes from the mechanism that maintains the pressure by doing
work on the piston. 
In the case of a compact star, this is the gravitational field, and the
transition to quark matter is driven by the higher gravitational binding energy
of the denser quark matter phase.
The work done per quark is
\beq
\label{compact:work}
\Delta W/N = -p dV/N = \frac{p_*}{N} 
  ( V_{NM} - V_{QM} )
\eeqn
which is just the negative of (\ref{compact:latH}): the work done by the
gravitational field provides exactly the energy needed to convert
the nuclear matter to quark matter.

It is clear that as long as the microscopic processes that convert
quark matter into nuclear matter occur on a much faster timescale
than the compression, no energy will be liberated. It is possible
for energy to be liberated if 
the compression to happens fast enough that the center
goes out of equilibrium and becomes
supercompressed, i.e.~metastable.

\subsection{Mixed phase vs sharp interface}

\begin{figure}[t]
\includegraphics[width=0.9\textwidth]{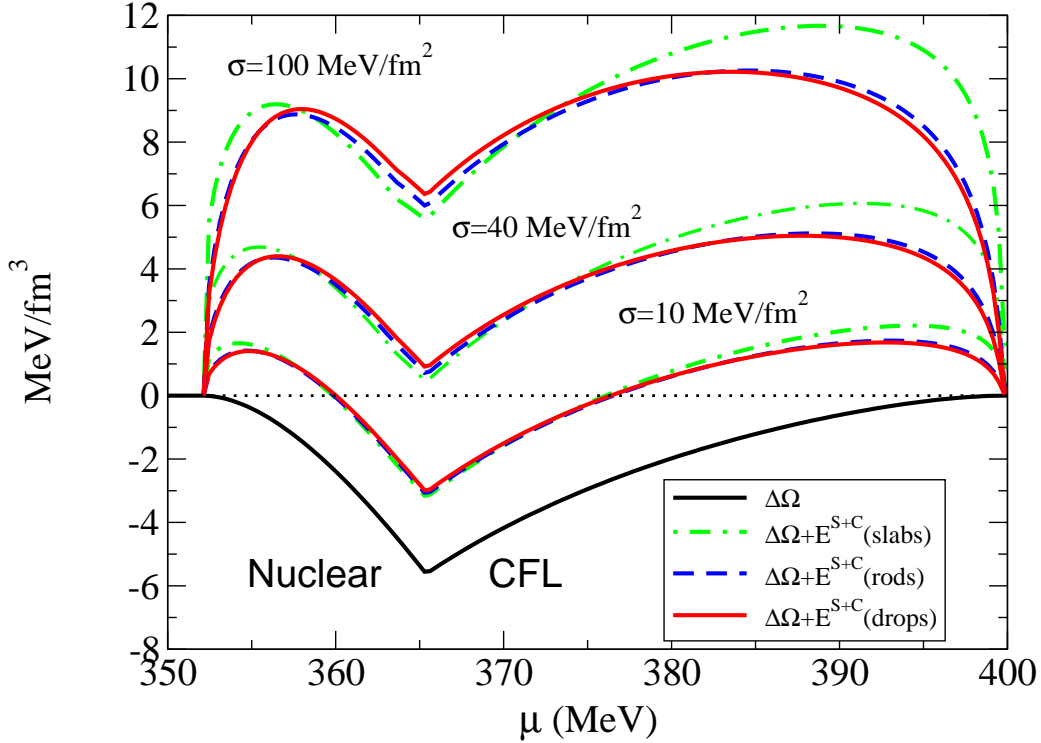}
\caption{The free energy difference between the mixed phase and the 
homogeneous neutral nuclear and CFL phases. In the lowest curve,
the surface and Coulomb energy costs of the mixed phase
are neglected, do the mixed phase always 
has the lower free energy. Other curves include surface
and Coulomb energy for different values of $\sigma_{\rm QCD}$ and 
different mixed phase geometry. As $\sigma_{\rm QCD}$ increases, 
the surface and Coulomb price paid by the mixed
phase increases.} 
\label{deltaomega}
\end{figure}

If the surface tension $\sigma_{QCD}$
and the electrostatic forces are ignored, then
a mixed phase is favored over a sharp interface.
\cite{Glendenning:1992vb,Prakash:1995uw,Glendenning:1995rd}.
If we  treat $\sigma_{\rm QCD}$ as 
an independent parameter, we can estimate the
surface and Coulomb energy cost of the 
mixed phase \cite{Glendenning:1995rd,interface}.
In  Fig.~\ref{deltaomega} we see how the competition between
sharp interface and mixed phase depends on the
$\sigma_{\rm QCD}$. The curves show the difference of free-energy
between various kinds of mixed phase and the sharp interface (which occurs
at $\mu=365~{\rm MeV}$, hence the kink there).
For any value of $\sigma_{\rm QCD}$, the mixed
phase progresses from drops to rods
to slabs of CFL matter
within nuclear matter to slabs to rods to drops of nuclear
matter within CFL matter.

For any given $\sigma_{\rm QCD}$, the mixed phase has
lower free energy than homogeneous neutral CFL or nuclear
matter wherever one of the curves in Fig.~\ref{deltaomega} 
for that $\sigma_{\rm QCD}$ 
is negative.  We see that much of the mixed phase
will survive if $\sigma_{\rm QCD}\simeq 10~{\rm MeV}/{\rm fm}^2$
while for  $\sigma_{\rm QCD} \gtrsim 40~{\rm MeV}/{\rm fm}^2$ 
the mixed phase is not favored for any $\mu$.
This means that if the QCD-scale surface tension
$\sigma_{\rm QCD} \gtrsim 40~{\rm MeV}/{\rm fm}^2$, the single
sharp interface with its attendant boundary layers,
described in previous sections, is free-energetically
favored over the mixed phase.

\section{Conclusions}
 
The reawakening of interest in the color superconducting nature
of quark matter has led us to appreciate that
the QCD phase diagram is much richer than previously thought.
I have outlined some of the structure in this paper.
There has naturally been much speculation about
observable consequences in compact star phenomenology.
In spite of the limited nature of our observational
knowledge of compact stars, there have been many suggestions,
and these are discussed in the review articles \cite{Reviews}
and in other contributions to these proceedings.
Although we are still in the early stages of such phenomenology,
we can take inspiration from high-temperature QCD, where
great progress has been made in overcoming similar obstacles.
At the same time as heavy-ion colliders map the high-temperature
region of the QCD phase diagram, we hope that astrophysical observations
and calculations
will complement it by filling in details of the
high-density region. 

{\bf Acknowledgements}. I am grateful to the organizers 
of Compact QCD, F.~Sannino and R.~Ouyed.
I thank J.~Bowers, S.~Hsu, K.~Rajagopal, S.~Reddy, D.~Son,
F.~Wilczek for
discussions and collaboration on the topics discussed here.

\end{document}